# Synthesis of $Fe_2O_3$ nanoparticles by new Sol-Gel method and their structural and magnetic characterizations

Shakeel Akbar[a], *S. K. Hasanain[a], Nasia Azmat[b] and M. Nadeem[c]

[a]Department of Physics, Quaid-I-Azam University Islamabad, Pakistan.
[b]Department of Chemistry, Quaid-I-Azam University Islamabad, Pakistan.
[c]Pakistan Institute of Nuclear Sciences and Technology, Nilore, Islamabad, Pakistan.

## ABSTRACT

$Fe_2O_3$ nanoparticles of different sizes ranging from 22 to 56nm were synthesized chemically by a modified sol-gel method. Pure alpha phase particles as well as particles with admixture of alpha and gamma phase were obtained and identified by x-ray and Mössbauer measurements. Different size and phase controlling parameters have been identified. The average size of the particles decreases with increased annealing temperature of the gel and decreases with the increase in the concentrations of the (citric acid). The annealing temperature affects the relative fractions of the two phases and consequently the magnetization of the particles. The magnetization of the particles ($M_s$) and coercivity ($H_c$) increase consistently with the increase in the annealing temperature. For the same relative amount of the two phases, the coercivity (Hc) and the moment of the particles increase with the decrease in the size of the particles, indicating the role of surface effects in the magnetic behaviour.

**Key words:** Nanoparticles, iron oxide; sol-gel; magnetization; Mössbauer.

* Corresponding author: skhasanain@qau.edu.pk
   Phone: +92-51-2872670     Fax: +92-51-2821397

# INTRODUCTION

From the viewpoint of the basic research, iron (III) oxide is a convenient compound for the general study of polymorphism and the magnetic and structural phase transitions of nanoparticles. The existence of amorphous $Fe_2O_3$ and four polymorphs (alpha, beta, gamma, epsilon) is well established [1]. The most frequent polymorphs, the hexagonal corundum structure "alpha" and cubic spinel structure "gamma", have been found in nature as hematite and maghemite minerals. The other polymorphs, the cubic bixbyite structure "beta" and orthorhombic structure "epsilon", as well as nanoparticles of all forms, have been synthesized and extensively investigated in recent years [1,2].

Gamma and epsilon type $Fe_2O_3$ are ferromagnetic; alpha $Fe_2O_3$ is a canted antiferromagnetic while beta type $Fe_2O_3$ is a paramagnetic material. The magnetic moment of bulk gamma type $Fe_2O_3$ is ~430emu/cc at room temperature while the magnetic moment of alpha type $Fe_2O_3$ is very small (~1emu/cc) [3] as compared to the gamma phase. $\alpha$-$Fe_2O_3$ (hematite) exhibits weak ferromagnetism above about 260K ™ [4], in addition to the antiferromagnetism below 960 K. This lower temperature transition is called the Morin transition. The observed weak ferromagnetism arises from the canting of the antiferromagnetically aligned spins above $T_M$ and this spin-canting angle is $\approx 5^o$ [5,6]. Nanoparticles of the alpha phase also exhibit this behaviour and the Morin temperature has been found to be strongly dependent on the size of the particles, generally decreasing with it and tending to disappear below a diameter of ~8nm for spherical particles [7].

On account of the attractive scientific and industrial applications of $\alpha$-$Fe_2O_3$ nanoparticles, novel methods for their synthesis and new approaches in their characterization have been reported in recent years. It has been reported [8,9] that different techniques of preparation lead to different phases or mixtures of phases and different degrees of size control. Furthermore the correlation of the preparation route with size and magnetic properties, such as the moment and coercivity, is still a subject of debate. In this paper we report the preparation of $Fe_2O_3$ nanoparticles ($\alpha$ and $\alpha+\gamma$) by a variation of the reported sol-gel method and correlate the properties of the obtained nanoparticles as functions of various preparation parameters.

## MATERIALS AND METHODS

Our objective was to prepare nanosize particles of hematite ($\alpha$-Fe$_2$O$_3$) by the sol-gel method and to identify the parameters that control their size. Previous studies have shown that this can be done by different methods. Nanocomposites of polypyrrole and iron oxide have been reported to be prepared using simultaneous gelation and polymerization processes. In one case [10] varied amounts of pyrrole monomer were added to a solution containing iron nitrate as precursor and 2-methoxy ethanol as solvent. Matijevic and Scheiner [11] synthesized hematite particles by dissolving ferric salts in the hydrochloric acid and heating at 100$^o$C. We have prepared these particles by a slightly modified method for greater simplicity. 200 ml (0.1M) of iron nitrate Fe(NO$_3$)$_3$. 9H$_2$O (Aldrich 98%) was used as a precursor solution, and was gelated by using 800 ml of mono hydrated citric acid (Aldrich 98%) solution (0.05 to 0.2M) as ligand molecules, and singly distilled water as the solvent. The iron solution was added to the citric acid solution drop wise with vigorous stirring. The solution was then heated to a temperature of 70$^o$C, while maintaining vigorous stirring until the gel was formed and the contained water was evaporated. The dried gel was annealed at temperatures ranging from 180-400$^o$C, typically yielding 1.6g of Fe$_2$O$_3$ ranging in size from 22-56 nm.

We used an x-ray diffractometer (JDX-11 Jeol) for the verification of the crystal structure and for the average size of the particles. A commercial vibrating sample magnetometer (VSM) was used for studying the magnetic properties and Mössbauer spectroscopy was used to complement the magnetic identification of the particles and the relative fractions of the alpha and gamma phases.

## RESULTS AND DISCUSSION

*Analysis and characterization of Sol-Gel prepared particles*

Two typical x-ray diffraction patterns that are representative of the general behaviour shall be discussed here. Fig.1 shows representative data for the samples annealed in the temperature range of 180 to 250$^o$C. This particular sample (sample #2) had been prepared by the previously described method where the molarity of both the iron nitrate and the citric acid was 0.1 while the obtained gel was annealed at 210°C. The major XRD peak was obtained at $2\theta \cong 33.2°$ while the second major peak was obtained at $2\theta \cong 35.7°$. For various samples belonging to this range of annealing temperature the relative intensity of these two major peaks lay between 70 to 80% The other observed peaks were at $2\theta = 41°$, $49.55°$, $54.25°$ and $62.55°$. While the 35.7° peak could be due to either of the alpha or gamma phases the relative intensities of the two major peaks and the other smaller peaks that are observed suggests strongly that alpha $Fe_2O_3$ is the major phase in this material. The lattice constants so obtained for this set of alpha $Fe_2O_3$ nanoparticles are $a = b = 5.0356$ Å and $c = 13.7489$ Å. All the remaining samples annealed below the temperature of 250°C gave a similar pattern, confirming that they are all predominantly of the $\alpha$-$Fe_2O_3$ phase.

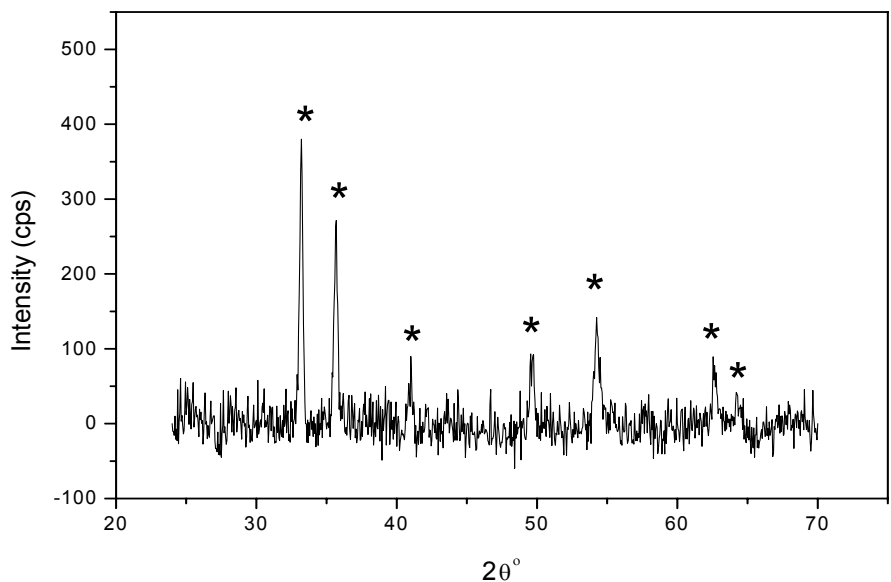

**Figure 1:** XRD pattern of the sample #2 annealed at 210°C with an average particle size of 33nm. (Symbol * represents the hematite peaks).

However the samples that were annealed at a higher temperature (400°C) gave somewhat different patterns with a suggestion of some of new peaks and a definite change in the ratios of the two major peaks at 33.2 and 35.7° as shown in Fig. 2. While the two major peaks are still at the same positions as in Fig. 1 but the identities of the highest and second highest peaks has been reversed. The 35.7° peak is now the most intense while the 33.2° peak is second highest with an intensity ratio of 0.8. In the pure gamma phase the relative ratios of the two major peaks $I_{33.2}/ I_{35.7}$ is much smaller. We believe that this is an indication that while both gamma and alphas phases are present in these samples prepared at higher temperature, the amount of the gamma phase is probably small and hence the smaller peaks are not visible. Hence while the contributions from the hematite are clearly visible at 33.2°, they overlap with those from the maghemite at 35.7° thus leading to the observed peak intensity. There is thus an indication that with the increase in annealing temperature (250-400°C) structural changes are occurring that allow the formation of the gamma phase along with the alpha phase. This pattern is confirmed decisively by the Mössbauer data discussed later.

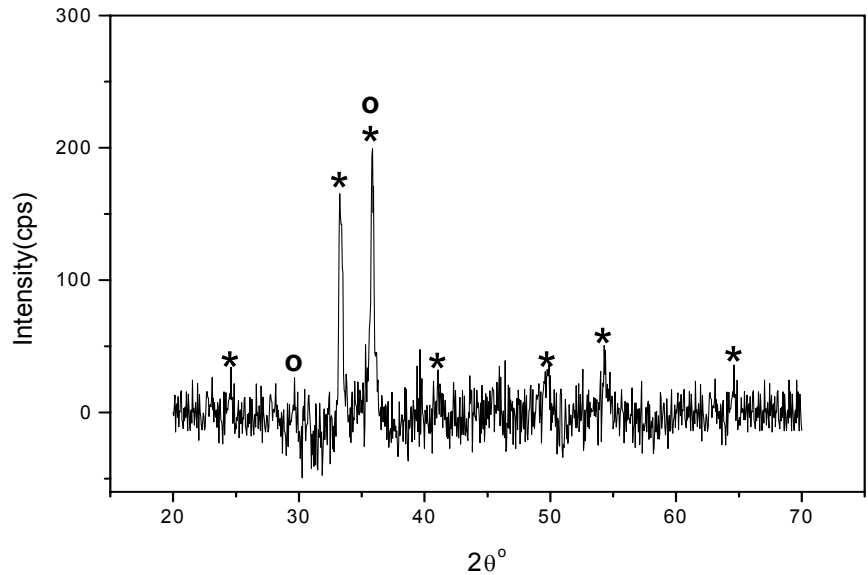

**Figure 2:** XRD pattern of the sample #8 annealed at 400°C with an average particle size of 33nm. (Symbols * and ° represent the peaks of hematite and maghemite respectively)

Average particle diameter D for different specimens was obtained from the main peaks using Scherrer's formula for the peak width broadening as a function of the size of the particles, $D = \dfrac{k\lambda}{\beta \cos(\theta)}$

Here λ is the x-ray wavelength (Cu K$_\alpha$ = 1.5418Å), k is the machine constant (0.916), β is the full width at half maximum of the peak and θ is the peak position [12]. Using the above method we obtained an average size of 33nm for particles of sample #2. The error estimated in the average size of the particles was about 2nm for a resolution $d\theta$ = 0.01º between the data points.

The detailed size variation with different parameters is discussed below.

*Size controlling parameters and their effects*

In the current sol-gel method we observed that the size of the particles depended on three parameters. These were: a-Dilution of the solutions, b-Concentration of the citric acid solution, c-Annealing temperature of the dried gel. In the following we discuss these effects briefly.

**A:** Mixtures with a higher ratio of the citric acid solution to iron nitrite are characterized by us as more dilute. Typically we mixed the iron solution in the citric acid solution in the ratio of 1:4. The more dilute the solution the less probable it is for the colloidals in the solution to interact with each other. Thus it appears that mixtures with higher dilution should result in smaller size particles. Comparing the sizes of the particles prepared from mixtures with the dilution ratios of 1:4 and 1:8 respectively, keeping all the other parameters of the reaction same, the size of the particles changes from 33nm to 27nm. While we have only these two sets of different dilutions, the trend here appears to be consistent with the expectation viz. smaller sizes for higher dilution.

**B:** The concentration of the ligand (citric acid) is the second factor affecting the size of the particles. With the increase in the concentration of the citric acid the number of citrate ions shielding the surface of the growing particles from agglomeration will also increase. Hence increasing concentration may tend to support the growth of smaller size particles. This trend is clearly visible in Table 1, which shows the particle size decreasing from 56±6 to 22±4nm as the concentration of citric acid changes from 0.05 to 0.2M (molarity).

**C:** The temperature at which the gel was dried is referred to in the literature [13] as the "annealing temperature". At this temperature the organic compounds decompose and other impurities are removed from the precursor i.e. the dried gel. We observe that this temperature has a very significant effect on the size of the final product, i.e. as we increase the annealing temperature from 180 to 250°C the size of the obtained particles decreases from 36 to 22nm. This is illustrated in Table 2. We suggest that this decrease in size may be due to the smaller chances of agglomeration for the particles emerging from the colloidal gel network, at higher average energies associated with the higher temperatures. It has been reported [13] that the variation of the annealing temperature can lead to an increase or decrease of the size of the particles depending on the concentration of the reactants and the particular route chosen for the preparation [13].

*Magnetic measurement on the different size particles*

In order to measure the magnetic properties of the particles we compacted them into pellets that were then cut into suitable sizes. Our measurements were focused on the determination of the magnetization of the particles as functions of their size and the different parameters of preparation. The magnetization, hysteresis and coercivity were determined. The DC measurements were made using a commercial Vibrating sample magnetometer (VSM). The hysteresis loops were studied at two different temperatures, 300 and 77K. Typical field cycling range was between ± 10kOe.

**Table 1**

**Effect of citric acid concentration on the nanoparticles properties.**

| Sample no. | Concentration (M) | Size (nm) | $M_s$ (emu/g) | $H_c$ (Oe) |
|---|---|---|---|---|
| 6 | 0.2 | 22 | 24.7 | 157 |
| 7 | 0.1 | 27 | 23.5 | 118 |
| 8 | 0.1* | 33 | 20.3 | 59 |
| 9 | 0.05 | 56 | 25.3 | 23 |

$M_s$ is the saturation magnetization and $H_c$ the coercivity. Note that all these samples were annealed at 400°C; all the samples were prepared at a dilution ratio of 1:4 except for the one shown starred* that had a ratio of 1:8.

**Table 2**
**Effect of annealing temperature on the nanoparticles properties.**

| Sample no. | Temperature (°C) | Size (nm) | $M_s$(emu/g) | $H_c$(Oe) |
|---|---|---|---|---|
| 1 | 180 | 36 | 0.3 | -- |
| 2 | 210 | 33 | 3.7 | 39 |
| 3 | 230 | 31 | 5.1 | 43 |
| 4 | 250 | 24 | 12 | 81 |
| 5 | 400 | 22 | 24.9 | 162 |

$M_s$ is the saturation magnetization and $H_c$ the coercivity. All samples were prepared at a citric acid concentration of 0.1M.

Fig. 3 shows the magnetization loop of sample #2 (33nm), prepared (annealed) at 210°C and as discussed above the x-ray data show the α phase to be present. This would be expected to show the weak ferromagnetism associated with this phase above 260K. The magnetization loop shows significant hysteresis ($H_c$ = 39Oe), and magnetic moment of 3.66emu/g (6.45emu/cc) at room temperature. At 77K the moment is slightly larger (3.72emu/g; 6.55emu/cc) while the coercivity increased from 39 to 35 7Oe. It is noticeable that the shape of the hysteresis loop is *constricted* at room temperature and becomes symmetric at 77K. Constricted loops are typically observed in materials with a mixture of a soft and hard magnetic phase [3]. Thus at room temperature the observed response could be due to the combination of the alpha and gamma phases, the gamma phase being a soft phase with a much higher moment and the alpha phase having a much higher coercivity but a much lower moment. At low temperature, on the other hand, the alpha phase is fully antiferromagnetic and the magnetization is essentially that of the gamma phase. The increase in coercivity at lower temperature could be due to the increase in the anisotropy energy of the gamma phase.

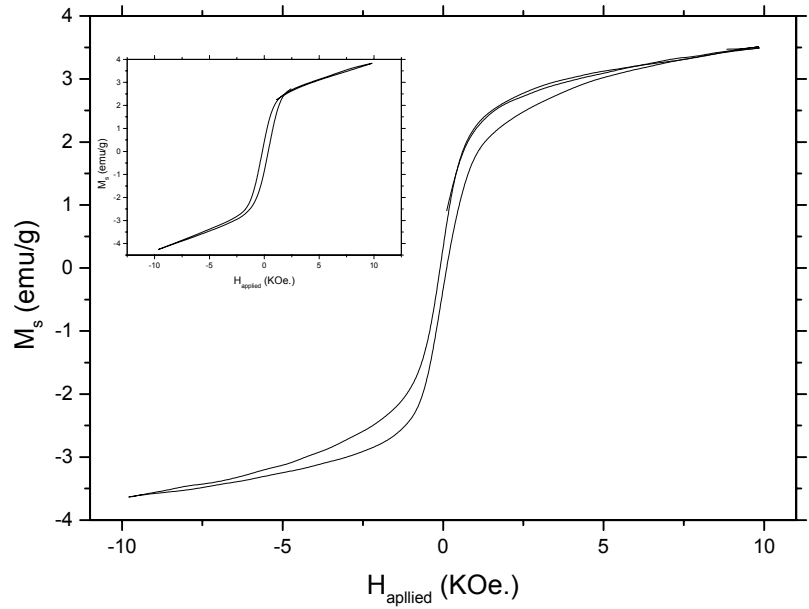

**Figure 3:** M-H loop of sample #2 (33nm) at 300K ($H_c$ = 39Oe). (Inset) M-H loop of sample #2 (33nm) at 77K ($H_c$ = 357Oe).

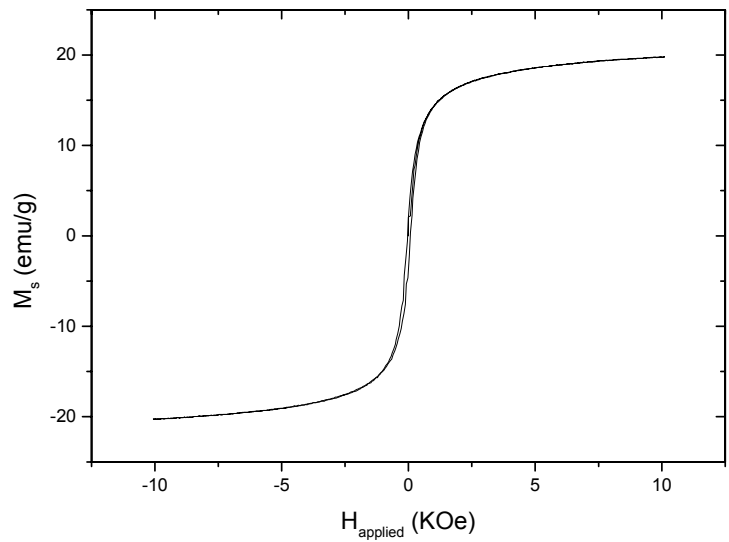

**Figure 4:** M-H loop of sample no. 8 annealed at $400^oC$.

The above magnetization values are to be compared to those for samples #1 and #12. Sample #1 was prepared with 0.1M concentrations of both citric acid and iron nitrate and was annealed at 180$^o$C. Sample #12 was prepared with a citric acid concentration of 0.2M and 0.1M iron nitrate concentration, and with aging of the dry precursors (gel) at 90$^o$C for about 16 hours in open atmosphere. Note that no higher temperature annealing was performed. These two samples had the lowest moments of all the samples, equal to 0.3emu/g (#1) and 0.89emu/g (#12) respectively, in very close agreement with several reports [4,14] for pure hematite nanoparticles. Mössbauer measurement on these samples showed peaks corresponding to the alpha phase only.

However for the other samples prepared by the variation of either of the parameters viz. concentration or temperature, we obtained magnetizations ranging between 3 and 25emu/g suggesting higher fraction of ferromagnetic phase. The coercivity of these particles lies in the range of 23 to 157Oe at room temperature. We find that the samples with higher moments exhibit an increased trend towards saturation (less high field slope of M(H) curve) than those having smaller moments. This is illustrated in Figures 3 and 4. This is consistent with the lower anisotropy of the gamma phase as compared to the alpha phase that is very hard to saturate. The particles with larger moments showed non-constricted loops presumably due to the domination of the gamma phase. Hence constricted loops only appear when a small amount of gamma phase (<5-6%) is present alongside the alpha phase. Variations in the particle moments with preparation parameters and sizes shall be discussed below.

*Discussion of magnetization*

We have already discussed that the magnetizations of our samples were generally much higher than expected for the pure hematite i.e. it should be ~ 1emu/cc while our values are generally in the range 6 to 90emu/cc, depending on different preparation parameters. We understand this as being due to the presence of the ferromagnetic gamma phase in small quantities in our particles. Note that due to the very large moment of γ-Fe$_2$O$_3$ (about 500 times) [3] as compared to α-Fe$_2$O$_3$, a very small amount of γ-Fe$_2$O$_3$ is sufficient to give very significant increase as compared to the moment of pure hematite.

For example we estimated the percentage of γ-Fe$_2$O$_3$ in the sample (#2) whose average magnetization was (3.66emu/g), using standard values of the moments of the two phases in the bulk i.e. 0.2emu/g for the alpha phase and 85emu/g for the γ-Fe$_2$O$_3$ phase. Hence the average moment for a sample with *x* and *1-x* fractions of α-Fe$_2$O$_3$ and γ-Fe$_2$O$_3$ respectively would be $M_s = (x)(M_{Fe_2O_3}) + (1-x)(M_{\gamma-Fe_2O_3})$

We obtain that about 4% of γ-Fe$_2$O$_3$ phase (1-x = 0.04) alongside the remaining alpha phase particles would be sufficient to give the observed signal. Thus we estimate, using bulk values of the moments of the two phases, that sample #2 having $M_s$=3.66emu/g contained about 96% of hematite and about 4% of γ-Fe$_2$O$_3$.

It is noticeable that for the samples annealed at 400°C the moment is even higher, lying between 70 to 90 emu/cc (20-25emu/g) depending on the molarity of the citric acid. The percentage of γ-Fe$_2$O$_3$ for one such case (sample #8, $M_s$=20.3emu/g) was estimated to be about 24%. The XRD pattern for these particles is shown in Fig. 2. where the peaks corresponding to γ-Fe$_2$O$_3$ phase are indicated. Mössbauer data on the same sample (#8) confirmed the presence of the gamma phase. Apparently the amount of γ-Fe$_2$O$_3$ becomes detectable via x-ray diffraction for compositions with higher annealing temperatures. Thus we conclude that in the samples annealed at the higher temperatures in this range the fraction of γ-Fe$_2$O$_3$ is higher.

*Effect of size on $M_s$ and $H_c$*

We can divide this discussion into two parts.
**1:** Particles whose size variation has been obtained by change of concentration of the citric acid (ligand molecules).
**2:** Particles whose size variation has been obtained by change of annealing temperature.

A: All these particles were annealed at 400°C. Here we do not find large changes in the magnetization with size of the nanoparticles in the range 22 to 56nm, though there does appear to be a slight decease in the moment with increasing size in the range 22-33nm. This data is shown in Table 1, where we observe that the moment lies between 20.3emu/g and 25.3emu/g. Thus the variations of the citric acid (ligand) concentration (for 400°C anneal) do not appear to change the relative amounts of the two phases to any great extent but do affect the sizes of the particles very significantly.

The variation of the coercivity with size is shown in Fig. 5 and is seen to be quite definitive where it consistently *increases* with the *decrease* in the particle size. The change is quite large ranging between 23 and 157Oe as the size decrease from 56 to 22nm.

B: We now consider the magnetic behaviour of particles whose size variation has been obtained by the variation of the annealing temperature, keeping the concentration of the ligand as constant. The variation in the coercivity and magnetic moment with annealing temperature is shown in Fig.6 and the dependence on the sizes so obtained is detailed in Table 2. We observe that the moment clearly decreases as the size of the particles increases. This suggests that along with the increase in the size of the particles the fraction of the ferromagnetic phase is decreasing. This trend is consistent with the reported behaviour [15] where the gamma phase is stable for particles of sizes d<30nm and the alpha phase for higher sizes. The presence of the gamma phase was clearly detectable in the sample with the highest moment, in this series. Thus our data show that while the alpha phase is predominant, its amount decreases consistently as the size decreases from 33 to 22nm.

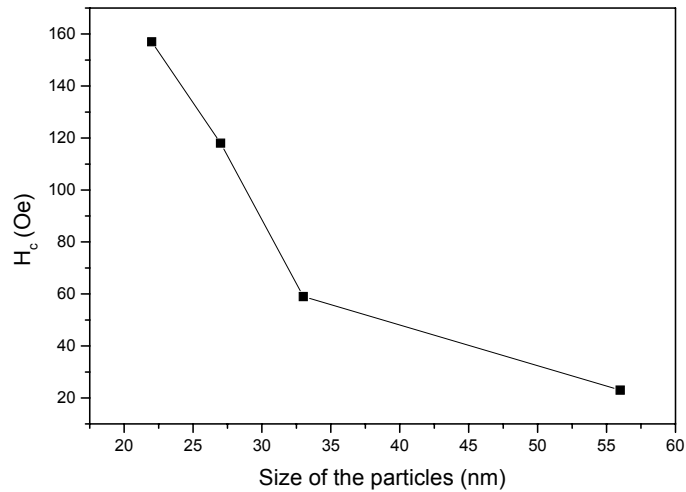

**Figure 5:** Effect particle size on the coercivity where particle size variation is obtained by varying the concentration of reactants. (see Table:1).

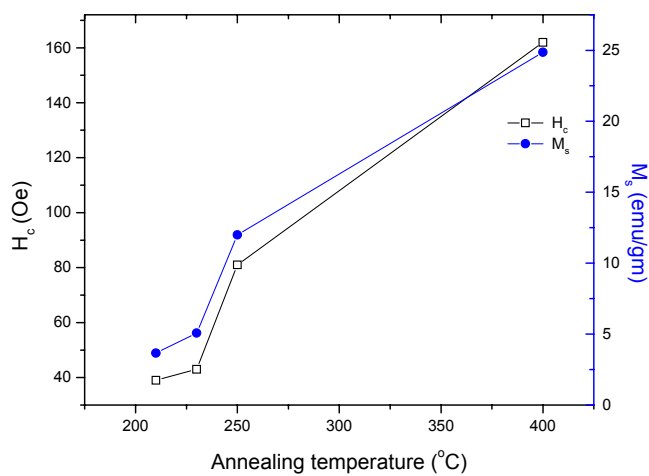

**Figure 6:** $M_s$ (emu/gm) and $H_c$ (Oe) vs. temperature, where size variation was obtained by the variation annealing temperature.(see table 2 for details)

It is apparent from the data of Table 2 that decreasing the size of the particles by the heat treatment at higher temperatures results not only in the increase in the magnetization but also in the coercivity, from 39 to 162Oe.

The increase of the coercivity with decrease in the size of the nanoparticles is thus common to both the methods of producing size variations. This trend of decrease of the coercivity with increasing size of single domain nanoparticles (above a critical size) has been reported by various authors [16-18]. Chen et al [17] explain this behaviour in nanoparticles in terms of an energy barrier ($\Delta E$) for the rotation of the moments that depends on the surface area of the particle, $\Delta E = K_s S$. Here $K_s$ is the surface anisotropy constant and $S$ the surface area of the particle. This has been shown to lead to a coercivity that decreases with the size of the particles as $H_c \sim 1/d$. Thus we understand that the observed increase in the coercivity of the particles with decreasing size reflects the increasing role of the surface in determining the anisotropy and hence the coercivity.

*Mössbauer studies*

The identification of the phases and their relative proportions obtained under different preparation conditions were estimated, as discussed above, by magnetometry. These were further checked by the use of Mössbauer measurements on several representative samples. The Mössbauer data was collected using a $^{57}$Co (Rh-matrix) source (initially activity of 25 mCi) in transmission geometry. The fitting of the data was carried out using a computer code Mos90 [19]. All spectra were taken at room temperature and were fitted assuming that all peaks are Lorentzian in shape.

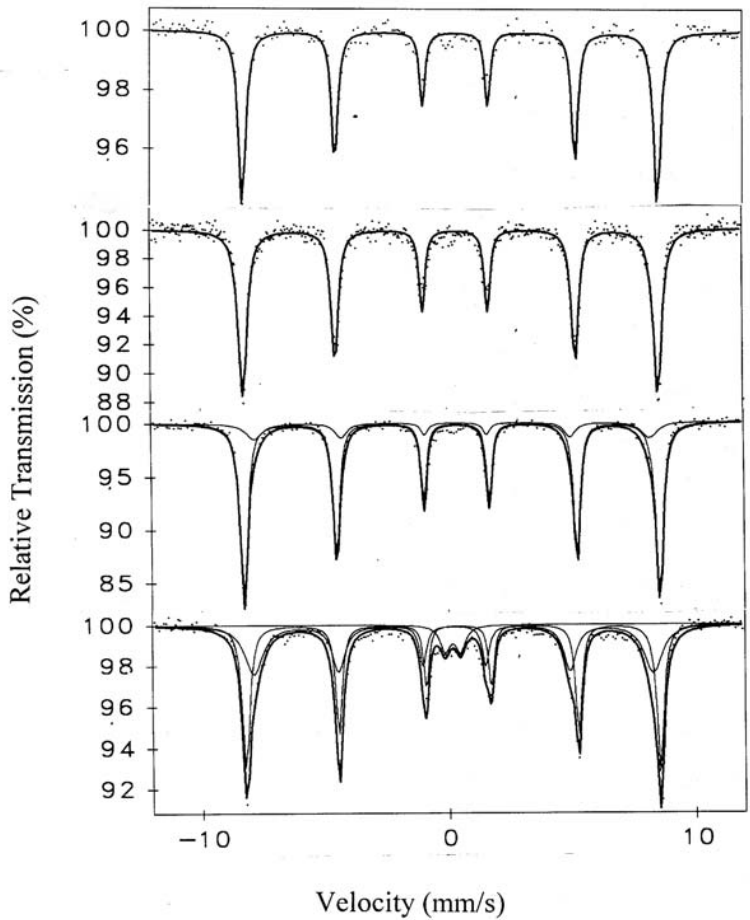

**Figure 7:** Mössbauer spectrum of samples 12, 1, 2 and 8 (from top to bottom) as discussed in the text. Best fit to the data is indicated.

The Mössbauer spectra for the samples #1 and #12 are given in Fig. 7. The parameters of the best-fitted spectra clearly indicate the presence *only* of α-Fe$_2$O$_3$. There is no signature of γ-Fe$_2$O$_3$ or any other phase of iron present in these samples. This is consistent with the magnetization measurements discussed earlier. However from the Mössbauer spectrum of samples #2 and #8 both the α-Fe$_2$O$_3$ and γ-Fe$_2$O$_3$ phases could be identified. The parameters of the best-fitted spectrum and the percentage of the two identified phases for each sample are reported in Table 3. The percentage of the phases is estimated from the fractional area under the respective peaks. For sample #2 the spectrum contained two subspectra viz. α-$Fe_2O_3$ (84% of the total sample) and γ-$Fe_2O_3$ (16%). In the case of sample #8 the main spectrum contained three subspectra viz. α-$Fe_2O_3$ (51% of the total sample); γ-$Fe_2O_3$ (43%) and a doublet that has been identified as γ-FeOOH (6%).

**Table 3**

**Mössbauer spectroscopy of different samples of nanoparticles prepared by the sol-gel method.**

| Sample no. | $H_{eff}$ KOe | QS (Δ) mm/s | IS (δ) mm/s | LW (Γ) mm/s | Area % | Phase |
|---|---|---|---|---|---|---|
| 12 | 522 | -0.213 | 0.37 | 0.35 | 100 | α-$Fe_2O_3$ |
| 1 | 522 | -0.206 | 0.37 | 0.42 | 100 | α-$Fe_2O_3$ |
| 2 | 523 | -0.209 | 0.37 | 0.35 | 84 | α-$Fe_2O_3$ |
|   | 498 | -0.206 | 0.37 | 0.76 | 16 | γ-$Fe_2O_3$ |
| 8 | 521 | -0.238 | 0.386 | 0.346 | 51 | α-$Fe_2O_3$ |
|   | 502 | -0.03 | 0.308 | 0.88 | 43 | γ-$Fe_2O_3$ |
|   | ** | -0.63 | 0.27 | 0.50 | 6 | γ-FeOOH |

Mössbauer parameters: Effective magnetic field ($H_{eff}$), quadrupole splitting (Δ), isomer shift (δ), line width (Γ) and area under the peak.

The results of magnetometry and Mössbauer are thus consistent in the qualitative sense regarding the relative fractions of the alpha and gamma phases obtained under different preparation conditions. We thus note that pure alpha phase as, confirmed by the Mössbauer; can be obtained by our processing route. The difference in the estimate of the quantity of gamma phase as determined by magnetization and Mössbauer techniques for samples #2 and #8 respectively, is not surprising given the fact that we used bulk magnetization values to estimate the fraction of the two phases from magnetometery.

## SUMMARY AND CONCLUSIONS

We observe that by using a modified sol-gel method we obtain best results for preparing alpha phase particles in the following two conditions:

**a:** Particles that were prepared at a citric acid concentration of 0.2M and 0.1M iron nitrate, with aging of the dry precursors (gel) at $90^{o}C$ for about 16 hours in open atmosphere. Note that in contrast to some other methods no higher temperature annealing at higher temperatures was required. (sample #12)

**b:** For samples annealed at high temperatures pure alpha phase particles were obtained for an annealing temperature of $180^{o}C$ with 0.1M concentration of both iron nitrate and citric acid.

We observe that variations in the concentration of the reactants are able to produce size variation without large changes in the quantity of the gamma phase. On the other hand variations of temperature of anneal in the range 180 to $400^{o}C$ produces similar changes in size but accompanied by large variations in the fraction of the gamma phase present.

We find the coercivity to be decreasing with increasing size of the particles independent of the way of varying the sizes or of the amount of gamma phase present. The decreasing trend of the coercivity with increasing size is attributed by us to the increased role of the surface anisotropy for smaller size particles. Large surface anisotropy values are quite common in nanoparticles systems. We also find that the presence of small traces of the gamma phase, undetectable by x-ray shows up in the magnetization measurements and leads to the development of constricted hysteresis loops. It is worth noting that the loop constriction vanishes at low temperatures presumably due to the absence of the weak ferromagnetism in the alpha phase.

After a series of the experiments we conclude that the particles size decreases as the annealing temperature is increased. But by varying this parameter we can only reduce the size of the particles up to a certain temperature range, effectively up to $250^{o}C$ where the size obtained is 24nm. Beyond this limit, instead of the expected decrease in the size, the fraction of γ-$Fe_2O_3$ starts to increase without very large change in the particle size. We note that in the samples where size control has been achieved by varying the temperature, the correlation of sizes with the gamma phase is very clear. Smaller size particles, obtained at higher temperatures, are more ferromagnetic and have more of the gamma phase. The correlation of size and the ferromagnetic behaviour is as discussed by Ayyub et al [15]. They report that for particle sizes smaller than d~30nm the ferromagnetic phase is stabilized. In our case increased temperature of anneal (210 <T<400$^o$C) not only produces smaller size particles but also helps stabilize the gamma phase. Our results are consistent with the above discussion with one reservation. The smaller size particles obtained by the variations of the concentration do not satisfy this criteria, with the moment being fairly independent of the size for the particles prepared at 400$^o$C. The size of the particles also shows a decreasing trend with increase in the concentration of the critic acid solution used as the ligand as well as the agent that controls the pH of the solution.

In summary we have successfully prepared nanoparticles, primarily of alpha $Fe_2O_3$, by the sol-gel technique and studied their magnetic characterization. The dependence of the particle sizes on different preparation parameters has been correlated and the dependence of the coercivity and the moment on the size have also been studied and reasonably explained.

Further experiments are required to understand how to vary the size beyond these limits and to be able to isolate the two phases in the oxide nanoparticles.

## ACKNOWLEDGEMENT

SKH acknowledges support from the Pakistan Science Foundation project no: PSF/ RES/ US/ NSF/ c-QU/ phy-18.